\documentclass[a4paper,fleqn,usenatbib]{mnras}

\usepackage{newtxtext,newtxmath}

\usepackage[T1]{fontenc}
\usepackage{ae,aecompl}

\usepackage{acronym}

\usepackage{graphicx}	
\usepackage{amsmath}	
\usepackage{tikz,xcolor,hyperref}

\definecolor{lime}{HTML}{A6CE39}
\DeclareRobustCommand{\orcidicon}{%
    \begin{tikzpicture}
    \draw[lime, fill=lime] (0,0) 
    circle [radius=0.16] 
    node[white] {{\fontfamily{qag}\selectfont \tiny ID}};
    \draw[white, fill=white] (-0.0625,0.095) 
    circle [radius=0.007];
    \end{tikzpicture}
    \hspace{-2mm}
}

\newcommand{\orcidSP}{\href{https://orcid.org/0000-0001-5579-9487}{\orcidicon}}
\newcommand{\orcidJUP}{\href{https://orcid.org/0000-0003-4291-2078}{\orcidicon}}
\newcommand{\orcidJW}{\href{https://orcid.org/0000-0002-2958-4738}{\orcidicon}}
\newcommand{\orcidSL}{\href{https://orcid.org/0000-0002-6948-0263}{\orcidicon}}
\newcommand{\orcidFW}{\href{https://orcid.org/0000-0002-0327-6585}{\orcidicon}}

\title[P-REx II. Off-line Performance on VLTI/GRAVITY]{Piston Reconstruction Experiment (P-REx) \\ II. Off-line Performance Evaluation with VLTI/GRAVITY}

\author[S. Perera et al.]{
Saavidra Perera,$^{1,2}$\thanks{E-mail: sperera@ucsd.edu}\orcidSP
Jörg-Uwe Pott,$^{1}$\orcidJUP
Julien Woillez,$^{3}$\orcidJW
Martin Kulas,$^{1}$ 
\newauthor Wolfgang Brandner,$^{1}$
Sylvestre Lacour,$^{4}$\orcidSL
Felix Widmann$^{1,5}$\orcidFW
\\
$^{1}$Max Planck Institute for Astronomy, Königstuhl 17, 69117 Heidelberg, Germany\\
$^{2}$University of California San Diego, 9500 Gilman Dr, La Jolla, CA 92093, United States of America\\
$^{3}$European Southern Observatory, Karl-Schwarzschild-Straße 2, 85748 Garching bei München, Germany \\
$^{4}$Observatoire de Paris, 61 Avenue de l'Observatoire, 75014 Paris, France \\
$^{5}$Max Planck Institute for Extraterrestrial Physics, Gießenbachstraße 1, 85748 Garching bei München, Germany \\
}

\date{Accepted XXX. Received YYY; in original form ZZZ}

\pubyear{2021}

\acrodef{AO}{adaptive optics}
\acrodef{CIAO}{Coudé Infrared Adaptive Optics}
\acrodef{DM}{deformable mirror}
\acrodef{FT}{fringe trackers}
\acrodef{LBTI}{Large Binocular Telescope Interferometer}
\acrodef{OPD}{optical path difference}
\acrodef{POL}{pseudo-open loop}
\acrodef{P-REx}{Piston Reconstruction Experiment}
\acrodef{PSD}{power spectral density}
\acrodef{rms}[rms]{root mean square}
\acrodef{r0}[r$_{0}$]{Fried's parameter}
\acrodef{TMT}{Thiry Meter Telescope}
\acrodef{TT}[TT]{tip/tilt}
\acrodef{VLT}{Very Large Telescope}
\acrodef{VLTI}{Very Large Telescope Interferometer}
\acrodef{WFS}{wavefront sensor}

\begin{document}

\label{firstpage}
\pagerange{\pageref{firstpage}--\pageref{lastpage}}
\maketitle

\begin{abstract}
For sensitive optical interferometry, it is crucial to control the evolution of the optical path difference (OPD) of the wavefront between the individual telescopes of the array. The OPD between a pair of telescopes is induced by differential optical properties such as atmospheric refraction, telescope alignment, etc. This has classically been measured using a fringe tracker that provides corrections to a piston actuator to account for this difference. An auxiliary method, known as the Piston Reconstruction Experiment (P-REx) has been developed to measure the OPD, or differential ‘piston’ of the wavefront, induced by the atmosphere at each telescope. Previously, this method was outlined and results obtained from LBT adaptive optics (AO) data for a single telescope aperture were presented. \ac{P-REx} has now been applied off-line to previously acquired VLT's GRAVITY CIAO wavefront sensing data to estimate the atmospheric OPD for the six VLTI baselines. Comparisons with the OPD obtained from the VLTI GRAVITY fringe tracker were made. The results indicate that the telescope and instrumental noise of the combined VLTI and GRAVITY systems dominate over the atmospheric turbulence contributions. However, good agreement between simulated and on-sky \ac{P-REx} data indicates that if the telescope and instrumental noise were reduced to atmospheric piston noise levels, \ac{P-REx} has the potential to reduce the OPD root mean square of piston turbulence by up to a factor of 10 for frequencies down to 1~Hz. In such conditions, \ac{P-REx} will assist in pushing the sensitivity limits of optical fringe tracking with long baseline interferometers.

\end{abstract}




\section{Introduction}
Temporal and spatial refractive index fluctuations of the atmosphere result in phase delays of different parts of the wavefront, thereby inducing speckling, movement and intensity fluctuations of an image. For interferometry, these induced phase delays result in unstable observed fringes. Since interferometry is reliant on accurately and precisely measuring the \ac{OPD} between each telescope (baseline),  correcting for these phase delays is critical for the overall performance of the instrument. 

Typically, this has been done by employing \ac{FT}.  Due to the short coherence time of the atmosphere (on the order of milliseconds), \ac{FT}s require bright targets to measure and stabilize the optical fringe position in real time. This limits the sky coverage and sensitivity of interferometry in the visible and infrared wavelength regime. 

To tackle this problem the new algorithm, \ac{P-REx}, can extend the integration time to be longer than the atmospheric coherence time, thus allowing observations of fainter targets \citep{Felix18}. \ac{P-REx} is an auxiliary method that uses only \ac{POL} measurement of the slope (gradient) of the wavefront phase from \ac{AO} data to reconstruct the temporal piston drift induced by the atmosphere. It can be used to stabilise the fringes over short timescales and extend the integration time of the fringe tracker, effectively increasing the coherence time of the system, thus improving sky coverage. The method has the advantage that the system can be easily implemented since all the required hardware (\ac{AO} and \ac{FT}) will already be available at the relevant interferometers. 

\ac{P-REx} relies on the assumption that the atmospheric profile is dominated by a single turbulent ground layer and follows \textit{Taylor`s Frozen Flow} hypothesis. In the following, we will demonstrate that this assumption is effectively met in approximately 2/3 of the data that was investigated for the study. With this assumption, \ac{P-REx} estimates that the piston drift is simply the product of the wind velocity and the \ac{TT} of the atmosphere. Therefore, an accurate real-time measurement of this ground layer wind velocity is crucial for the implementation of \ac{P-REx}. 
Previous work on \ac{P-REx} has included end-to-end simulations \citep{Felix18,Pott16} and preliminary off-line tests of on-sky data \citep{Felix18}. These tests were made on a single 8~m class telescope by applying a virtual aperture in order to imitate two telescopes. This paper discusses the performance of \ac{P-REx} off-line on the \ac{VLTI}'s GRAVITY system, which operates in the K-band. Comparisons of \ac{OPD} estimates by \ac{P-REx} from GRAVITY's \ac{CIAO} \ac{WFS}  \citep{Scheithauer16} data and GRAVITY's \ac{FT} \citep{GRAVITY-firstlight} data are presented. 

The \ac{P-REx} concept will be outlined in section~\ref{sec:method}. section~\ref{sec:TTest} presents \ac{TT} estimates from  \ac{CIAO} \ac{WFS} with comparisons to simulation. Section~\ref{sec:velocity} outlines the modified wind velocity estimation method, as well as estimates from on-sky \ac{CIAO} \ac{WFS} data.  In section~\ref{sec:prex_results} off-line comparisons of \ac{P-REx} to the  \ac{VLTI}'s GRAVITY \ac{FT} system will be presented. Please note that all simulated data presented in this paper was acquired using  SOAPY \citep{reeves16} and based on the \ac{VLT}'s \ac{CIAO} system (see table \ref{tab:ciao-properties} in section~\ref{sec:method}).

\section{P-REx Method} \label{sec:method}
 The \ac{P-REx} method is described here in brief for completeness. For a detailed discussion please refer to \citet{Felix18}. \ac{P-REx} uses \ac{POL} slopes to estimate the atmospheric piston drift over a single telescope. In a closed loop \ac{AO} system the \ac{POL} slopes must be computed to reconstruct the full wavefront information. This is achieved by summing the measured residual slopes from the \ac{WFS} and the reconstructed slopes that would be given if the \ac{DM} was flat \citep{Basden19},
\begin{equation}
\label{eq:fov}
S_{n}^{pol} = S_{n}^{res} + IM \cdot V_{n-k} \, ,
\end{equation}
\noindent where $S_{n}^{res}$ is the WFS measurements for frame $n$, $IM$ is the interaction matrix, $V$ is the voltage applied to the \ac{DM} and $k$ is the frame delay for the application of the voltages. The piston drift over a single telescope can be estimated by taking the product of the wind velocity and the gradient optical phase induced by the atmospheric turbulence,
\begin{equation}\label{eq:DP}
    \Delta P = [tip \cdot v_{x} + tilt \cdot v_{y}] \cdot \Delta t \, ,
\end{equation} 
\noindent where $v$ is the wind speed in the $x$ and $y$ direction and $\Delta t$ is the time over which the piston variation is measured. 

Equation \ref{eq:DP} is based on the following assumptions: (1) tip/tilt are the dominant optical aberration modes of atmospheric turbulence, (2) the atmospheric turbulence profile is dominated by a single layer at the ground level and (3) the temporal evolution of this single layer is well described by Taylor’s frozen flow model, i.e. ‘boiling’ of the wavefront is not relevant to the evolution of the piston on the relevant timescales (seconds).

By obtaining $\Delta P$ at each individual aperture of an interferometer, the difference in the piston evolution, i.e. the \ac{OPD} caused by the atmosphere, can be determined for each baseline
\begin{equation}
    OPD(T) = \sum_{t=0}^{T} \Delta P_{1}(t) - \sum_{t=0}^{T} \Delta P_{2}(t)  \, ,
\end{equation}
\noindent where the subscript 1 and 2 indicates the two telescopes in a single baseline.

\begin{figure}
    \centering
    \includegraphics[width=1.\columnwidth]{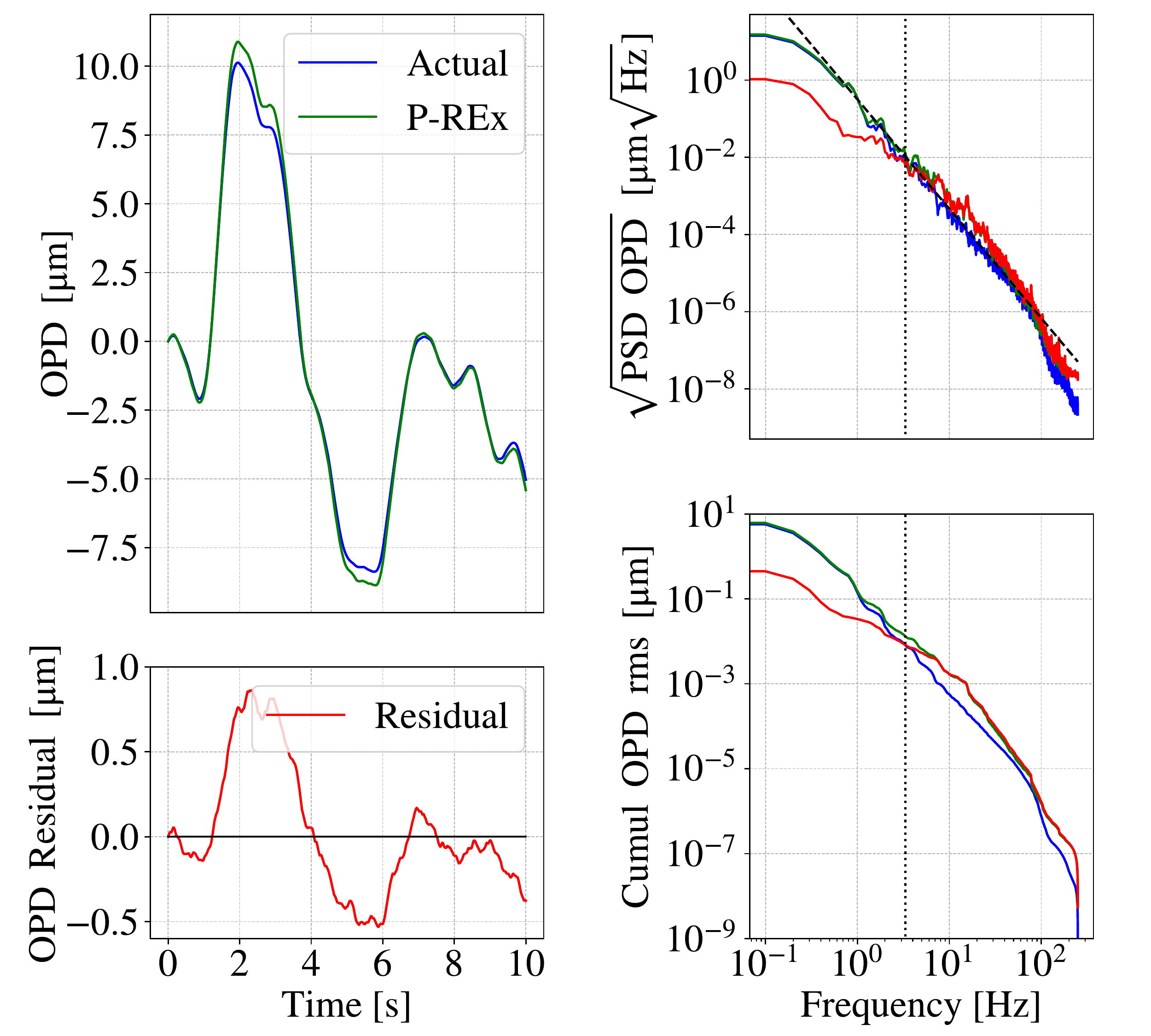}
    \caption{Example simulation of the actual OPD value taken from the phase screens (blue line) and \ac{P-REx} estimated atmospheric OPD (green line) for a single turbulent layer, of \mbox{$r_{0}$ = 10 cm,} traversing two 8 m telescopes with a speed of 10~m~s$^{-1}$. The red lines indicate the difference (residual) of the two values, the dashed black line indicates the theoretical atmospheric piston trend \citep{Conan95} and the black dotted lines indicates the turnover frequency where, for higher frequencies, the rms of the residual becomes greater than the rms of the actual OPD. (Left) example time series, (right-top) the averaged square root PSD of the OPD and (right-bottom) the averaged reverse cumulative of the OPD.} 
    \label{fig:simulated_prex_10s}
\end{figure}

\begin{table}
	\centering
	\caption{Properties of the GRAVITY CIAO WFS system}
	\label{tab:ciao-properties}
	\begin{tabular}{ll} 
		\hline
		Keyword & Description\\
		\hline
		Telescope diameter & 8.2 m \\
		$\#$ Subapertures & 9 x 9  \\
		FoV & 2\," \\
		Pixel scale & 0.5\,"/pix \\
		Wavelength & 1950\,nm \\
		Frame Rate & 500\,Hz \\
		DMs & Tip/Tilt Mirror +\\ 
		&  High Order DM (HODM) \\
		$\#$ HODM actuators & 60\\

		\hline
	\end{tabular}
\end{table}

Figure~\ref{fig:simulated_prex_10s} shows a simulated example of how the actual and \ac{P-REx} estimated \ac{OPD} compares for a single turbulent layer traversing two 8.2 m telescopes, based on the GRAVITY's CIAO system (see \mbox{table \ref{tab:ciao-properties}).} The actual values were taken directly from the phase values of the simulated atmosphere. In this example the simulated atmosphere is represented by a single turbulent layer at the ground where the \ac{r0} is 10 cm (at 500 nm), wind speed 10~m~s$^{-1}$ and direction 45~$^{\circ}$ relative to the axes of the \ac{WFS}. The \ac{P-REx} estimates were derived from the simulated \ac{POL} slopes generated when observing this atmosphere. In this case the standard deviation of the residual is 0.38~$\mu$m. Figure~\ref{fig:simulated_prex_10s} shows for frequencies less than 1~Hz, the \ac{PSD} of the residual is more than a factor of ten less, with a \ac{rms} of approximately a factor of 5 less. Simulations show that the \ac{rms} of the \ac{P-REx} estimated OPD residual is dependent on the wind speed and \ac{r0} of the atmospheric turbulent layer. 

\citet{Conan95} show that the square root of the \ac{PSD} of the atmospheric piston follows a frequency$^{-17/6}$ dependence at high frequencies. The simulation shows that \ac{P-REx} follows this trend until a turnover at approximately 3.3~Hz where, for higher frequencies, \ac{P-REx} measures an excess of power in the OPD variations. As a result, \ac{P-REx} can no longer reduce the power of the atmospheric OPD at these higher frequencies.  This overshoot is in part due to the limited spatial sampling of the \ac{WFS}. Simulations showed that the turnover frequency increased when the spatial sampling (i.e. number of sub-apertures) of the WFS was increased. In addition, \citet{Conan95} show how the power of higher order aberrations start to dominate at higher frequencies. An extension to this work would be to explore additional aberrations. However, the current aim is to first assess the \ac{TT} aberrations. This excess power could be filtered out in a real-time \ac{P-REx} controller, although its amplitude does not significantly add to the total OPD power.

\ac{P-REx} could be used as a ‘tweeter’ in a two-stage fringe stabilization scheme. At very low frequencies (below 1~Hz) a classical fringe tracker with a slow duty cycle would act as a ‘woofer’ to remove small inaccuracies of \ac{P-REx} that accumulate over many cycles.

Since \ac{P-REx} is a cumulative algorithm, small errors can result in a large deviation from the true differential piston evolution. Therefore, accurate measurement of the \ac{TT} and the ground layer wind velocity is imperative. 

\section{Estimation of atmospheric Tip/Tilt}
\label{sec:TTest}
\begin{figure}
    \centering
    \includegraphics[width=1.\columnwidth]{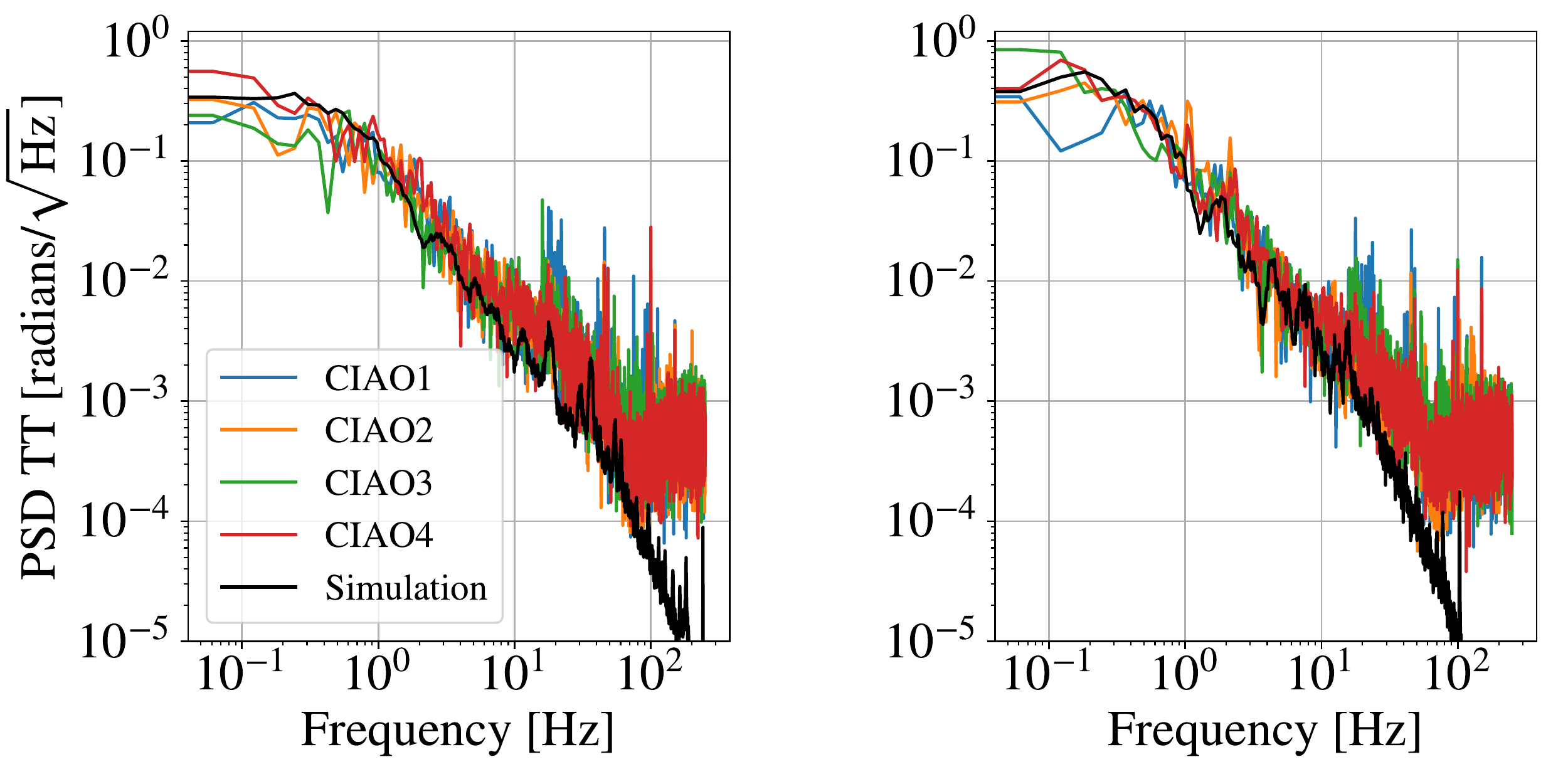}
    \caption{Two examples of the averaged PSD of the estimated TT from simulation (black) and CIAO WFS results, under different estimated atmospheric conditions: (left) r$_{0}$ = 0.12 m and wind \mbox{velocity = 16.6~m~s$^{-1}$}, (right) r$_{0}$ = 0.1~m and wind \mbox{velocity = 10~m~s$^{-1}$}. } 
    \label{fig:TTexample}
\end{figure}

Estimating the atmospheric \ac{TT} is necessary for the \ac{P-REx} algorithim. The global \ac{TT} is estimated by averaging the \ac{POL} slopes over the full aperture for each exposure. Whilst it is not possible to know if the measured \ac{TT} is accurate, a useful cross-check is to compare the \ac{PSD} of the averaged \ac{TT} estimates from simulated and on-sky CIAO data. Figure~\ref{fig:TTexample} shows an example of this, where the atmospheric properties of the simulations were based on the CIAO report of \ac{r0} and P-REx's measurement of wind velocity. Figure~\ref{fig:TTexample} shows a good agreement between the simulated and on-sky measurement of the PSD of the TT, for frequencies lower than 10~Hz. This indicates that the atmospheric model used in simulations were appropriate and that what is being measured is the atmosphere and not additional instrumental contributions for lower frequencies.

\section{Estimation of the turbulence wind velocity} \label{sec:velocity}  

\begin{figure}
    \centering
    \includegraphics[width=1.\columnwidth]{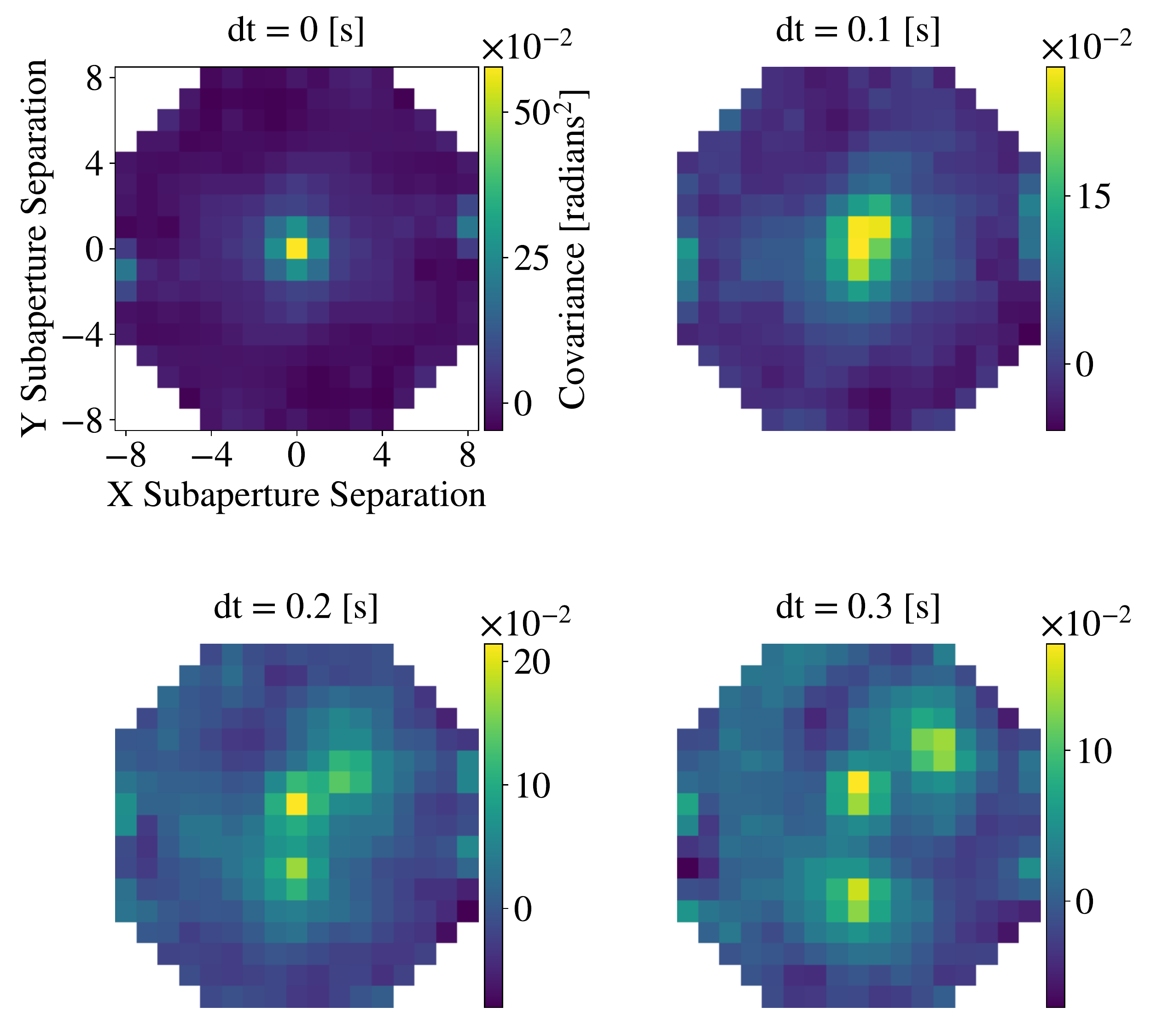}
    \caption{Example spatio-temporal covariance map from slopes of simulated three layer turbulence profile with wind speeds 5, 10 and 15~m~s$^{-1}$, wind direction 0, 45 and 180 $^{\circ}$  and relative strengths 0.4, 0.3 and 0.3,  respectively. } 
    \label{fig:simulated_3layers}
\end{figure}

Assuming frozen flow, the pattern of the \ac{POL} slopes traverse the telescope aperture at the wind speed of the corresponding turbulent layer. Hence the dominant turbulent layer wind velocity can be acquired from the spatio-temporal auto-covariance of the x and y \ac{POL} slopes,\begin{equation}
T_{\delta i, \delta j, \delta t} = \langle C_{i,j,t} C'_{i',j',t'} \rangle \, ,
\end{equation}
where $C$ and $C'$ are the slopes at subaperture position $[i,j]$ and $[i',j']$, and at time $t$ and $t'$ respectively. The spatial offset between the subapertures, in units of the subaperture diameter, are given as $\delta i$ and $\delta j$. This method has been employed by many optical atmospheric turbulence profilers such as SLODAR \citep{Wilson02} and SCIDAR \citep{Sheperd13}. By calculating this for every possible separation for a given temporal offset, $\delta t = t - t'$, a covariance map can be created. If $\delta t = 0$ then a peak will appear in the centre of the map, which corresponds to the superposition of all of the turbulent layers. However, for increasing values of $\delta t$,  a peak corresponding to each layer will be offset by an amount related to the distance and direction that the turbulent layer has travelled in $\delta t$ \citep{Osborn10, PereraThesis, Perera20}. 

Figure~\ref{fig:simulated_3layers} illustrates this with the averaged x, y slope covariance maps with increasing temporal offset values. The atmospheric profile was represented by a three layers at altitudes 0, 5 and 10 km, wind speeds 5, 10 and 15~m~s$^{-1}$, wind direction 0, 45 and 180 $^{\circ}$  and relative strengths 0.4, 0.3 and 0.3,  respectively. It should be noted that this is not a typical atmospheric turbulence profile, it was purely chosen to act as a visual aid. The intensity of each peak is dependent on the strength of the corresponding turbulent layer. 

\begin{figure}
    \centering
    \includegraphics[width=1.\columnwidth]{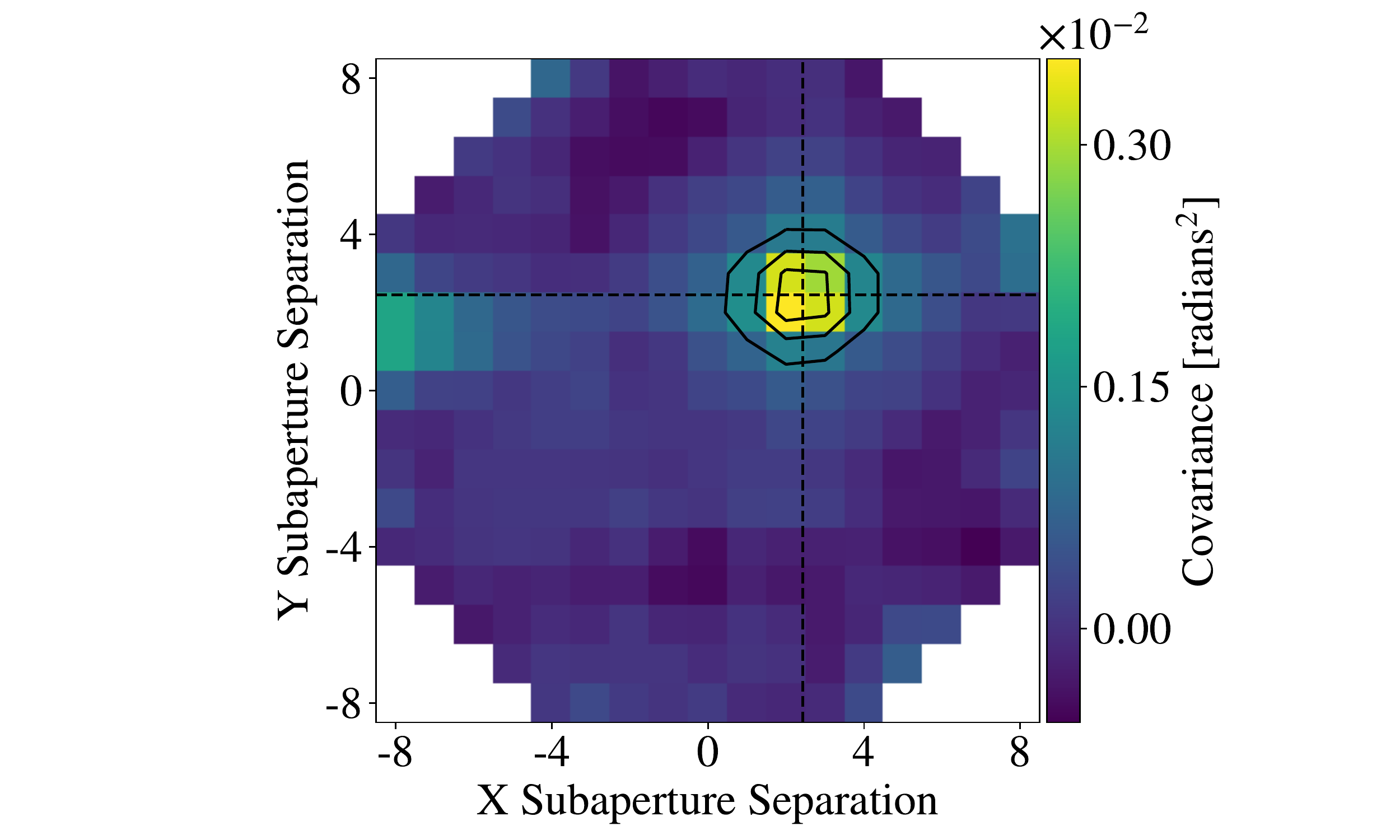}
    \caption{Example of Gaussian fit to the peak covariance value to identify it's position. The input wind velocity parameters were 10~m~s$^{-1}$ at 45 $^{\circ}$, the output was 10.2 $\pm$ 0.2~m~s$^{-1}$ at 45.3 $\pm$ 0.5 $^{\circ}$. Figure taken from \citep{Perera20}. }
    \label{fig:gauss_position}
\end{figure}

In order to obtain the wind velocity of the dominant layer, the position of the strongest peak is found and a 2D Gaussian fit is employed for a measurement of  sub-pixel precision \citep{Felix18}, see figure~\ref{fig:gauss_position}. Due to the effects of slowly developing dome turbulence it is important that the peak is at least a single subaperture separation away from the centre. Due to statistical error, it should not be too close to the edge. The edges of the covariance pattern correspond to larger sub-aperture separations, for which there are a smaller number of contributing sub-aperture pairs. Hence, averaging of the covariance is reduced and the statistical error is larger.

\subsection{Measurement Frequency \& Averaging}
\begin{figure}
    \centering
    \includegraphics[width=1.\columnwidth]{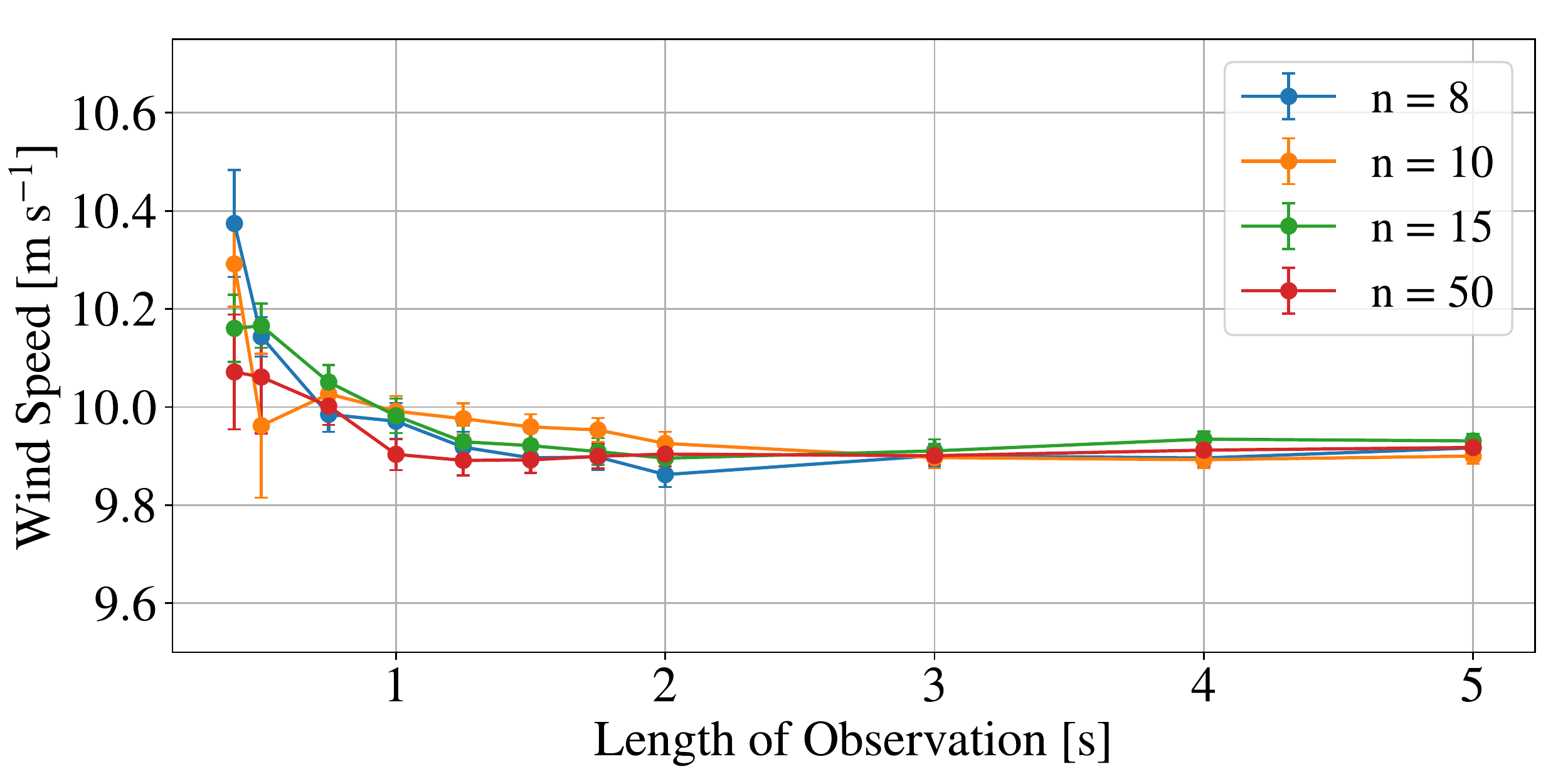}
    \caption{Estimated wind speed, from simulated data, for increasing length of observations. Each line represents results for $n$ number of photons per subaperture for a given exposure time of 2 ms. The input velocity for the single turbulence layer was 10~m~s$^{-1}$. The error is the standard error found from independent repetitions of the simulation. }     
    \label{fig:shotnoise}
\end{figure}

Here the requirements for accurate estimation of the wind velocity are explored, in terms of the sampling frequency and length of observations, i.e. the number of consecutive seconds in which data was taken at a given frame rate. Figure~\ref{fig:shotnoise} shows the mean estimated wind speed measurements for simulated results in the presence of shot noise, for $n$ number of photons per subaperture, for an exposure time of 2~ms and with frame rate of 500~Hz. The input wind speed for this simulation was 10~m~s$^{-1}$. For longer observations the error decreases and estimated wind velocity value converges to \mbox{9.91 $\pm$ 0.01}~m~s$^{-1}$. In addition to the statistical error, there appears to be a small offset bias of 0.1~m~s$^{-1}$ between the input and converged output values. This is consistent for different wind speeds. It originates from the finite spatial resolution of the covariance map, i.e. in the size of the \ac{WFS} subapertures. This offset decreases for increasing resolution. 

For short observations, on average, there is tendency to overestimate the wind speed. For these observation lengths there is less temporal averaging, resulting in more statistical uncertainty, particularly for larger WFS subaperture separations. This biases the measurements to slightly higher wind speeds. The minimum temporal sampling rate required for an accurate wind velocity measurement for $\delta t$ = 0.25 s is $\sim$1 - 2 s. Observations less than 0.75 s tend to overestimate the wind speed. The figure also shows that for as few as $n$ = 8 photons, when an appropriate observation length is used, \ac{P-REx} can accurately estimate the wind speed.

\begin{figure}
    \centering
    \includegraphics[width=1.\columnwidth]{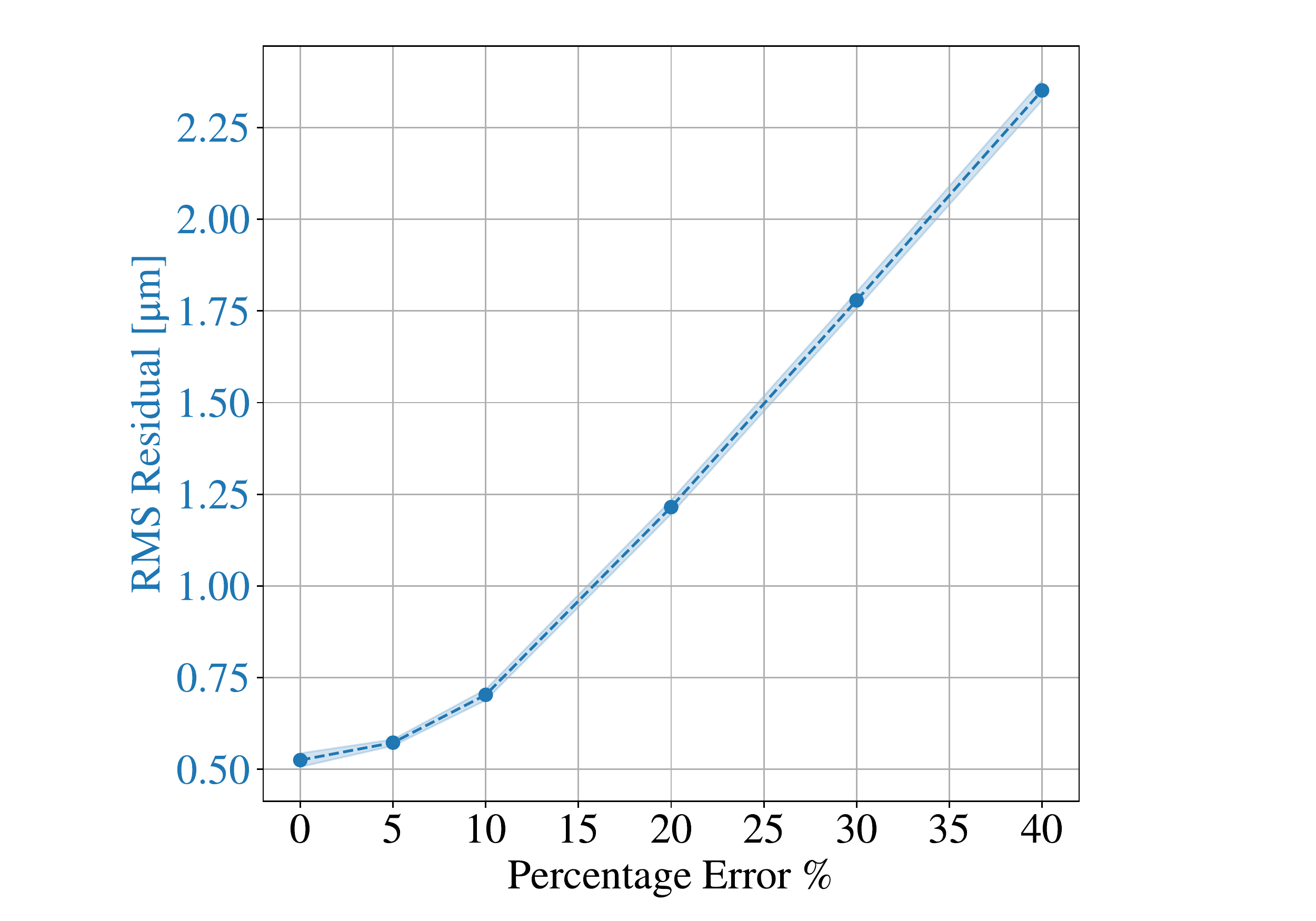}
    \caption{Relationship between the OPD rms and the percentage error in the wind speed measurement.} 
    \label{fig:mean_corr_res}
\end{figure}

Figure~\ref{fig:mean_corr_res} shows how the atmospheric OPD rms increases with the percentage error in the measurement of the wind velocity. For example, the rms increases by less than a factor of two in going from a 10~$\%$ error to a 20~$\%$ error of the wind velocity measurement.

\begin{figure}
    \centering
    \includegraphics[width=1.\columnwidth]{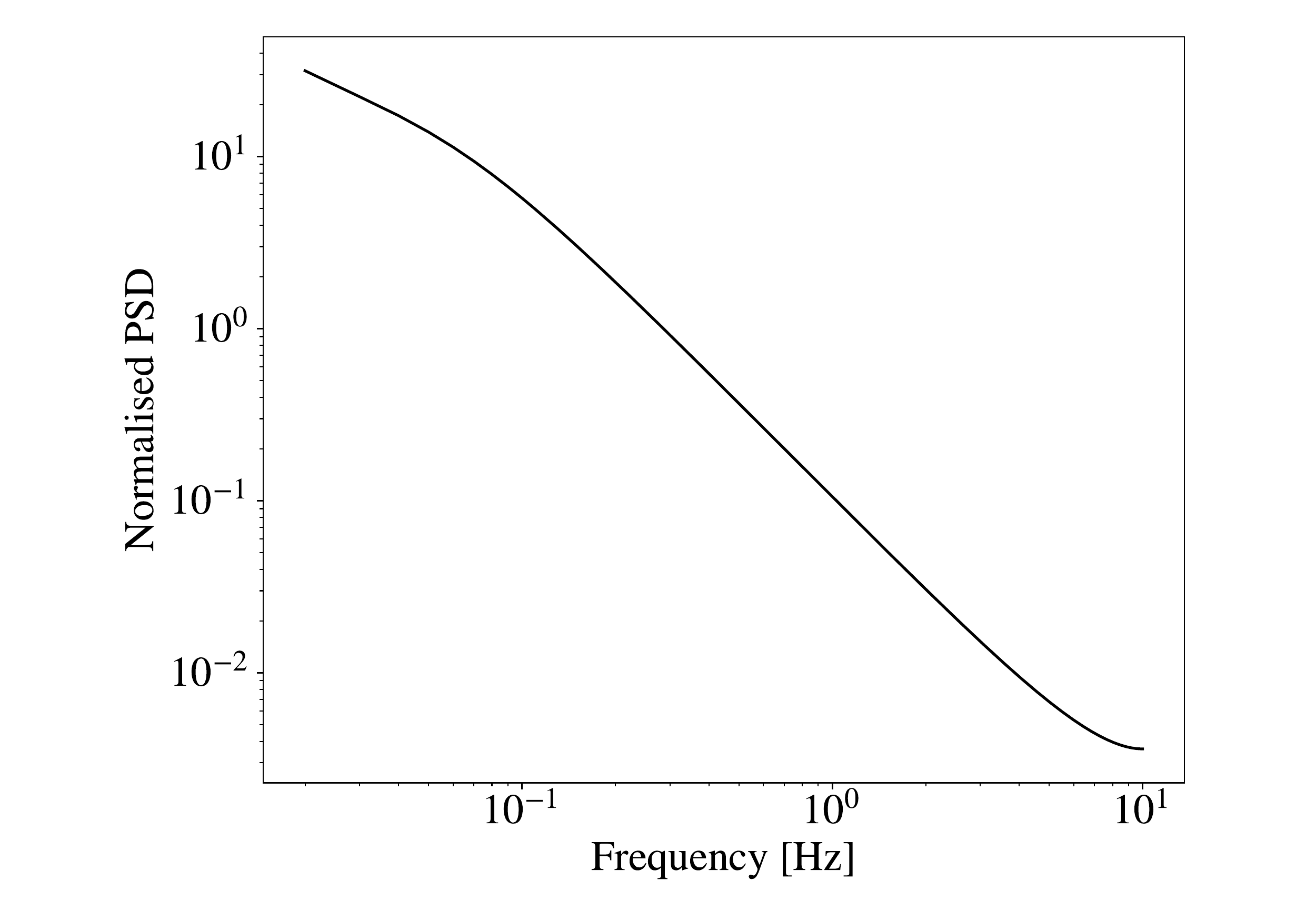}
    \caption{ARTFIMA model used to describe the average wind speed normalised PSD (dimensionless)  measured at the TMT site at Mauna Kea \citep{vanKooten19}.  } 
    \label{fig:vankootenplot}
\end{figure}

The random fluctuations of the wind speed during the observation time will affect the estimate of the wind speed. Here the size of the resulting uncertainty and its effect is estimated. \citet{vanKooten19} acquired the average wind speed \ac{PSD} from wind data taken from the \ac{TMT} site testing campaign at Mauna Kea, over a period of three years (2006-2008).  Figure~\ref{fig:vankootenplot} show the ARTFIMA model \citep{Meerschaert14} they fitted to describe the average normalised \ac{PSD} of the measured wind speed. The \ac{PSD} covered the frequency range 0.002 - 10~Hz, with power clearly increasing at lower frequencies. According to the ARTFIMA model $\sim$95 $\%$ of the power occurs for frequencies less than 1~Hz, indicating that most variation happens over longer time scales. A sampling rate decrease from 2~Hz to 0.1~Hz results in over a 200 $\%$ increase in \ac{rms}. This is significantly higher compared to a sampling rate change from 2~Hz to 1~Hz, which results in a 37 $\%$ increase in \ac{rms}. Therefore, it is reasonable for \ac{P-REx} to measure the wind speed to a temporal resolution of this order of magnitude ($\sim$ 1~Hz), since typically, no significant faster variations are expected. This was confirmed in the analysis of the provided CIAO data, which showed only a minimal difference between wind estimates when changing observation length between \mbox{2 - 10 s}. In addition, as shown by figure~\ref{fig:mean_corr_res}, small percentage errors in the wind speed estimate do not greatly impact the results. Since GRAVITY operates in the K-band (2.2~$\mu$m), it therefore requires the percentage error of the wind speed to be less than $\sim$35~$\%$.

The wind velocity estimate algorithm was run 10,000 times on 2~seconds of consecutive simulated AO data, with a data acquisition frame rate of 500~Hz. This was to assess the time taken to run the algorithm, from opening the saved \ac{WFS} data to producing a wind velocity value. The median time taken was 280~ms with a standard deviation of 27~ms. This estimate was performed on a 16 GB MacBook Pro 2.7~GHz Intel Core i7-8559U with 4 cores. Although this is a relatively long time compared to the \ac{AO} frame rate, this is close to a factor of 7 faster than the acquisition of the  2~seconds of \ac{AO} data required to estimate the wind velocity. Therefore, this is adequate for real time processing.


\subsection{VLTI-CIAO Data Wind Velocity Estimates}
The \ac{P-REx} wind velocity estimation algorithm was applied off-line to CIAO \ac{WFS} data to verify the assumption that the atmospheric profile is dominated by a single turbulent layer at the ground. The algorithm was applied to 576 CIAO \ac{WFS} datasets, for all four Unit Telescopes (UT), spanning 24 nights over April to August 2018.  

\begin{figure}
\begin{center}
\begin{tabular}{c} 
\includegraphics[width=1.\columnwidth]{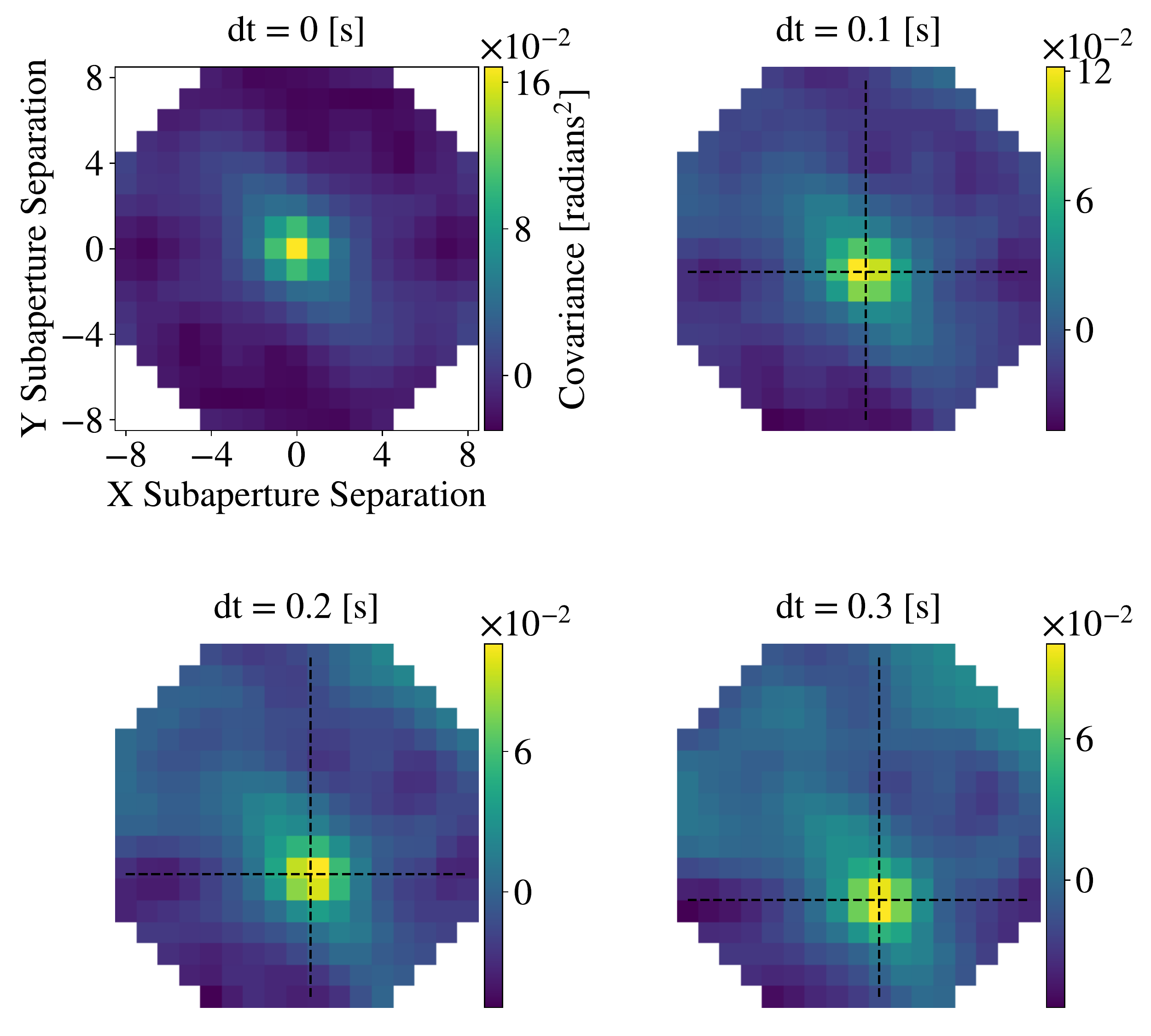}
\end{tabular}
\end{center}
\caption{Spatio-temporal auto-covariance map generated from UT 1's POL CIAO data, for increasing temporal offsets of $\delta t$. There is a clear single dominant turbulent layer that traverses the telescope with a wind speed of $\sim$10.5~m~s$^{-1}$ at 74.4 $^{\circ}$.} \label{fig:ciao_varyingdt}
\end{figure}

\begin{figure}
\begin{center}
\begin{tabular}{c} 
\includegraphics[width=1.\columnwidth]{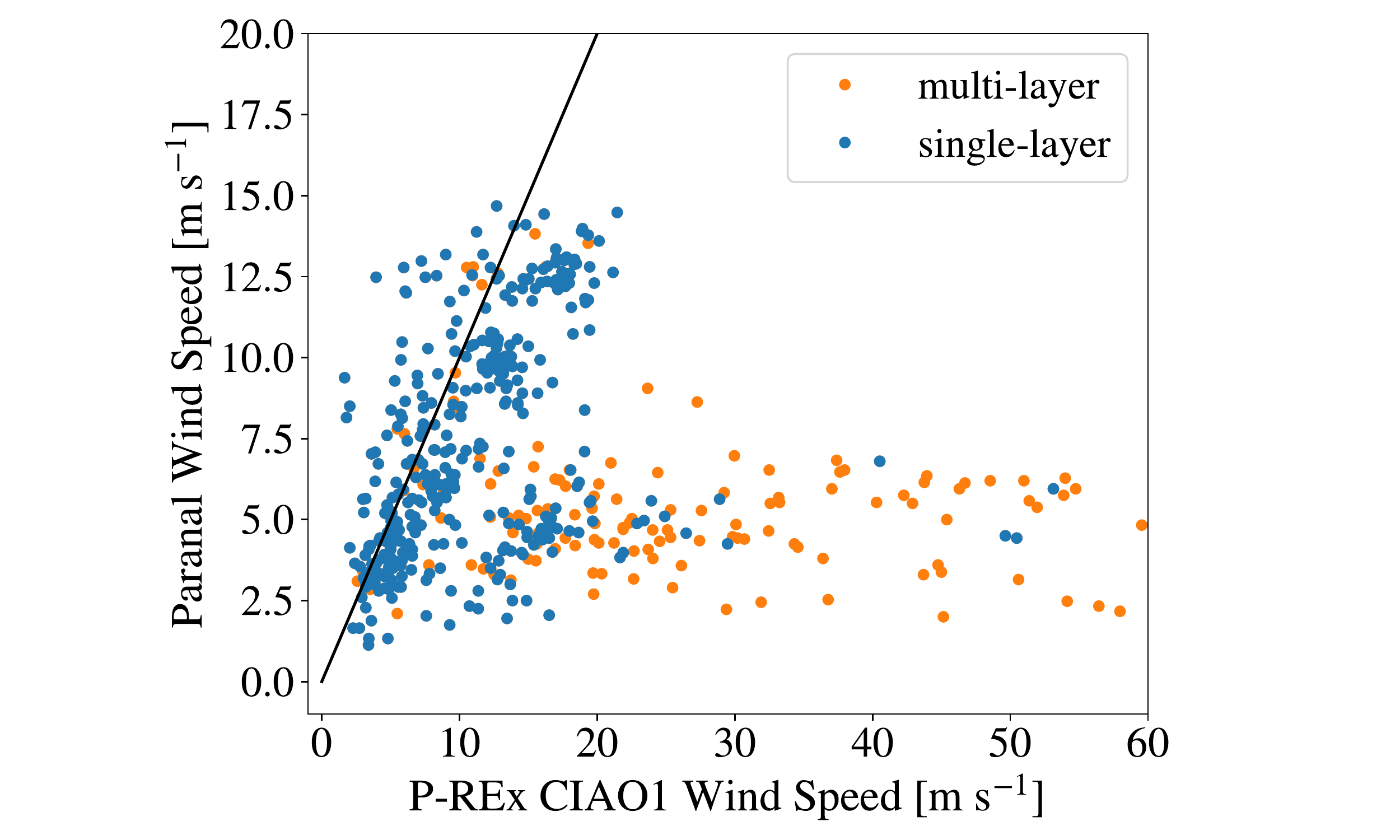}
\end{tabular}
\end{center}
\caption{Comparison of wind speed estimates acquired from CIAO WFS data for UT1 and Paranal's wind speed measurement. The blue markers indicate when a single dominant turbulent layer is observed in the covariance maps and orange markers when there are multiple turbulent layers with significant strength. The black solid line is x = y.   
} \label{fig:paranal_ciao_comparisons}
\end{figure}

Figure~\ref{fig:ciao_varyingdt} shows a typical covariance map for increasing values of $\delta t$, from on-sky CIAO \ac{WFS} data. In this example, there is a clear single dominant layer traversing the telescope pupil. \citet{Osborn10} presented SCIDAR statistical data showing that the ground layer is usually the strongest individual layer. In addition, \citet{Felix18} show through simulating a typical atmospheric profile of Paranal that the the ground would dominate such that only a single layer would be observed. This prior research, as well as results presented in this section (see figure~\ref{fig:paranal_ciao_comparisons}), indicate this dominant layer will most likely be located at the ground. The presence of a single strong turbulent layer indicates that the assumption described in section~\ref{sec:method} can be used.

\begin{figure}
\begin{center}
\begin{tabular}{c} 
\includegraphics[width=1.\columnwidth]{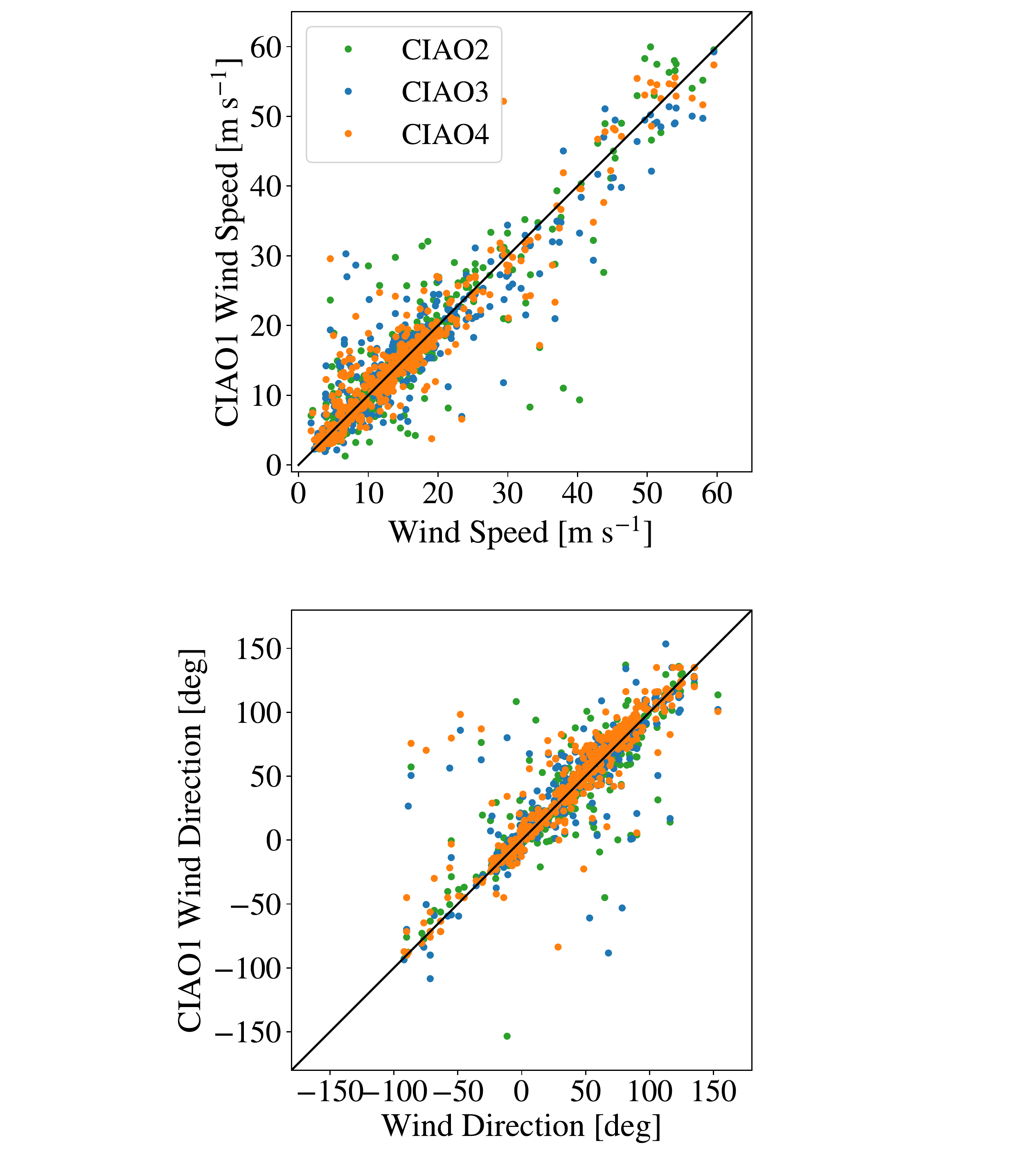}
\end{tabular}
\end{center}
\caption{Comparison of wind (top) speed and (bottom) direction estimates acquired from CIAO WFS data for UT2, UT3 and U4 against UT1 } \label{fig:ciao_comparisons}
\end{figure}

Figure~\ref{fig:ciao_comparisons} compares the wind velocities estimated from each CIAO system. There is a clear correlation between the wind velocities observed by each telescope. This indicates that the ground layer turbulence observed by each telescope is the same and that over this telescope separation wind velocities do not differ substantially. In addition, this tells us that there are no systematic noise biases associated with any of the individual CIAO systems. 

\begin{figure}
\begin{center}
\begin{tabular}{c} 
\includegraphics[width=.9\columnwidth]{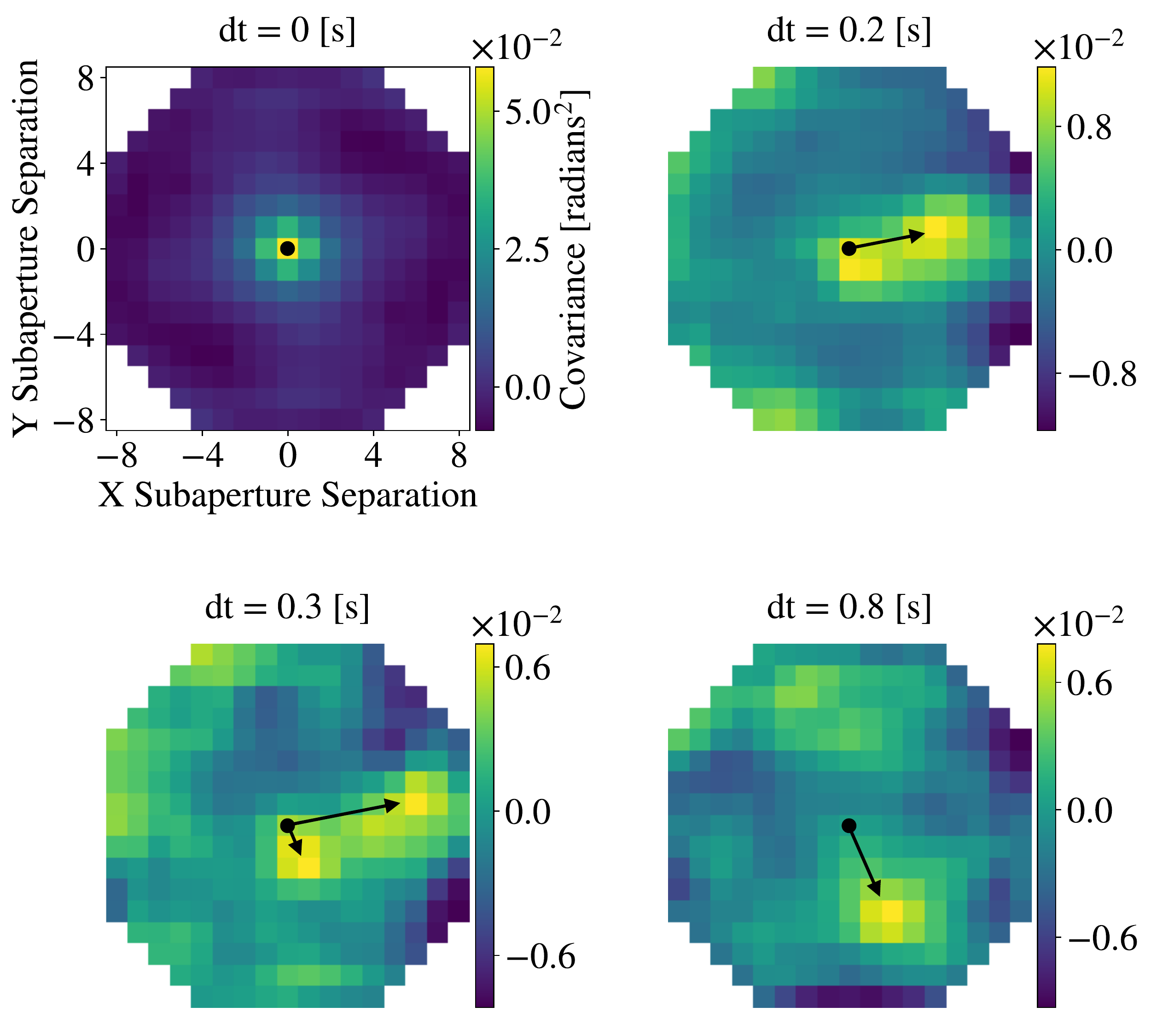}
\end{tabular}
\end{center}
\caption{Spatio-temporal auto-covariance map generated from POL CIAO1 data, for increasing temporal offsets of dt. Multiple turbulent layers can be observed with wind speeds of 17.0~m~s$^{-1}$ at 83.2 $^{\circ}$ and  4.27~m~s$^{-1}$ at 154.6 $^{\circ}$, where N=0 and E=90. } \label{fig:ciao_varyingdt_multi}
\end{figure}

Approximately 15~$\%$ of the data show dominant turbulent layers with wind speed greater than 20~m~s$^{-1}$ and at times reaching 50~m~s$^{-1}$. This is of course an unrealistic estimate for a ground turbulent layer, since the VLT closes their domes when local wind speeds exceed 18~m~s$^{-1}$. On closer inspection, for $\sim$82~$\%$ of these cases multiple strong turbulent layers were present. Figure~\ref{fig:ciao_varyingdt_multi} shows an example of a spatio-temporal covariance map with a strong multi-layer profile. Approximately 24~$\%$ of the data shows that there are multiple strong turbulent layers and \mbox{$\sim$80~$\%$} of these occurred on 8 of the 24 nights. The data indicated that both the wind velocity and relative strength of the ground layer can change significantly over the course of a night. 

Figure~\ref{fig:paranal_ciao_comparisons} shows the comparison between the \ac{P-REx} wind estimates from UT1 CIAO data and the Paranal on-site wind speed measurements. The Paranal measurements were retrieved from the file headers of the CIAO data, which are estimated by the Paranal Astronomical Site Monitoring (ASM)\footnote{http://archive.eso.org/cms/eso-data/ambient-conditions/paranal-ambient-query-forms.html}. The figure shows that for a single dominant turbulent layer, like that shown in figure~\ref{fig:ciao_varyingdt}, there is a good correlation between the \ac{P-REx} estimates and the on-site wind speed measurement.

P-REx requires specific atmospheric conditions, therefore 32~\% of the CIAO data provided was considered unusable. Of those discarded datasets, $\sim$72~\% showed signs of multiple, strong turbulent layers. These cannot be handled separately by the current version of P-REx, which assumes a dominant single turbulent layer. The remaining \mbox{$\sim$28~\%} of the discarded data is suspected to be unusable due to very slow turbulent layers, for which the wind velocity could not be accurately measured, or were exceptionally noisy data. Approximately 70~$\%$ of the discarded datasets were recorded over 7 nights. Based on this dataset, we expect that \ac{P-REx} could be straight forwardly applied to $\sim$2/3 of the typical AO-suitable nights on Cerro Paranal. 

\section{P-REx VLTI Results} \label{sec:prex_results}
The prediction performance of the \ac{P-REx} algorithm was tested \hbox{off-line} on recorded CIAO WFS data. Comparisons between concurrent \ac{P-REx} OPD estimates and VLTI GRAVITY \ac{FT} OPD values were compared for 17 datasets spanning 6 nights over April to August 2018. See \citep{Lacour19} for information on the FT POL OPD. 

\begin{figure*}
	\includegraphics[width=\textwidth]{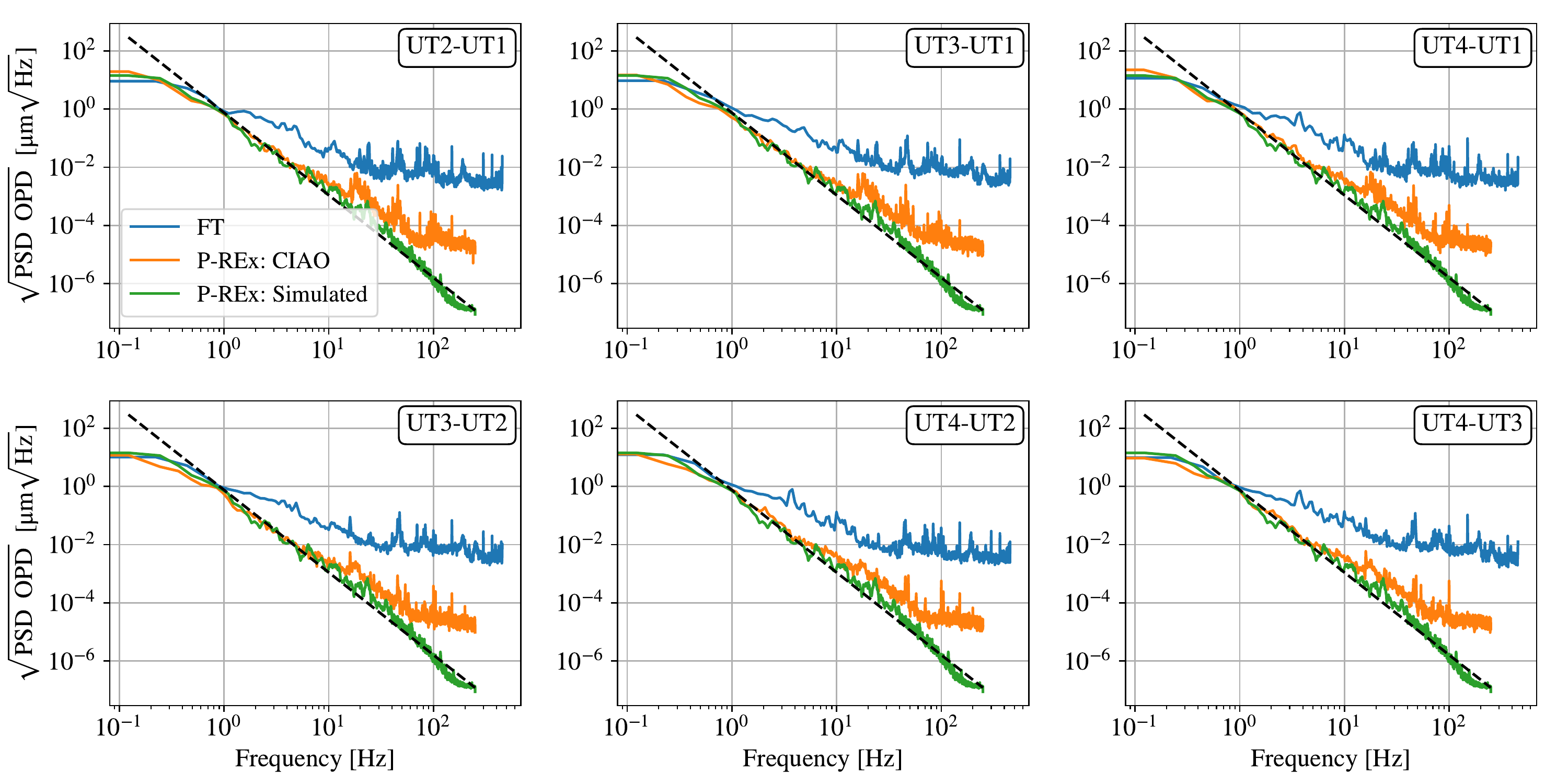}
    \caption{Example PSD of the OPD estimated by the GRAVITY FT (blue), \ac{P-REx} applied to CIAO data (orange), \ac{P-REx} applied to simulated data for the same atmospheric conditions (green) and the theoretical atmospheric piston trend $\propto$~frequency$^{-17/6}$.  The data was taken on the 27th June 2018.}
    \label{fig:psd-results}
\end{figure*}

\begin{figure*}
	\includegraphics[width=\textwidth]{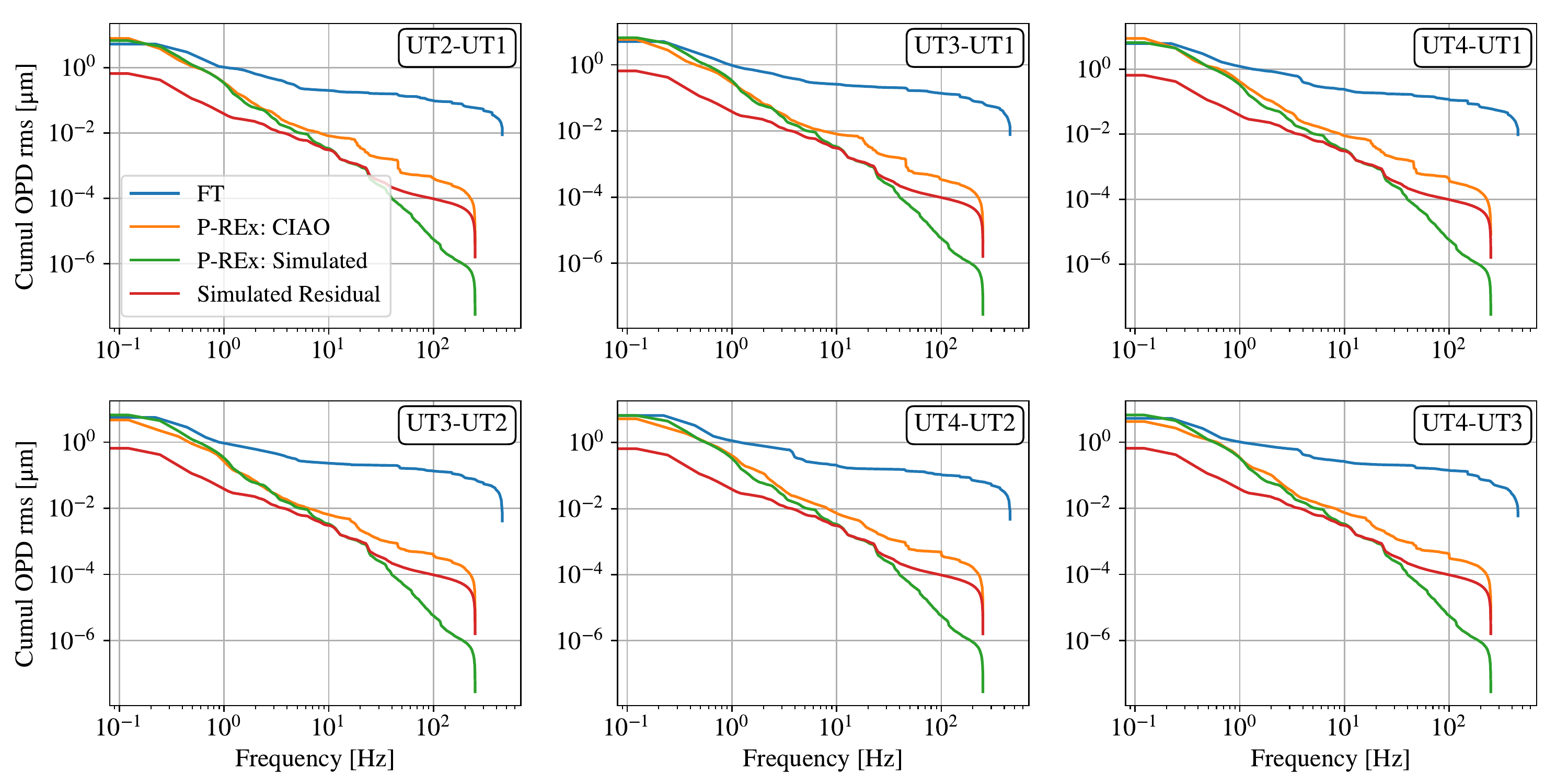}
    \caption{Example reverse cumulative of the OPD PSD estimated by the GRAVITY FT (blue), \ac{P-REx} applied to CIAO data (orange) and \ac{P-REx} applied to simulated data (green). Additionally, the simulated expected residual (red) is plotted, i.e. the difference between the actual OPD and the \ac{P-REx} estimated OPD found in simulation. The data was taken on the 27th June 2018. }
    \label{fig:cumulative-results}
\end{figure*}

Figure~\ref{fig:psd-results} shows an example PSD of the OPD estimates from the GRAVITY FT, \ac{P-REx} applied to CIAO WFS data and \ac{P-REx} applied to simulated WFS data, for each of the 6 baselines. In this example, the simulated atmosphere was based on the atmospheric parameters retrieved from the file header of the CIAO data (estimated from the Paranal ASM) and the wind speed estimated by P-REx. It can be seen that for frequencies less than 10~Hz there is a strong agreement in the power between the simulated and on-sky \ac{P-REx} estimates. This indicates that \ac{P-REx} can estimate the atmospheric OPD between telescope baselines as seen by the AO system. However, the \ac{FT} OPD estimates have a significant excess of power for frequencies higher than 1~Hz, above that of the atmospheric piston variation. In addition, the agreement of the theoretical piston trend and \ac{P-REx} further indicates that there are other non-atmospheric contributions to the GRAVITY FT measurements. Figure~\ref{fig:cumulative-results} highlights the difference in the \ac{rms} between the FT OPD and \ac{P-REx} OPD, by showing the reverse cumulative sum of the OPD PSD for the example used in figure~\ref{fig:psd-results}. The potential that \ac{P-REx} could reach is shown by the simulated residual. This is the difference between the piston calculated directly by the phase of the simulated atmosphere and \ac{P-REx} estimated value from the simulated AO system observing that atmosphere. It shows that for this specific ground layer profile the \ac{rms} can be reduced by up to a factor of $\sim$9 for frequencies higher than 1~Hz. Figures \ref{fig:psd-results} and \ref{fig:cumulative-results} show the results for a single dataset taken on the 27th June 2018. This example is representative of the general trends found in the remaining datasets.  The average \ac{rms} at 1~Hz for the 17 GRAVITY FT datasets and baselines was 1.4~$\mu$m, whereas, for the concurrent \ac{P-REx} estimates this was 0.7~$\mu$m. The benchmark of 1~Hz was qualitatively chosen since at very low frequencies, \ac{P-REx} prediction errors will build up and the FT would not benefit from applying such predictions at frequencies much lower than 1~Hz.

\begin{figure*}
	\includegraphics[width=\textwidth]{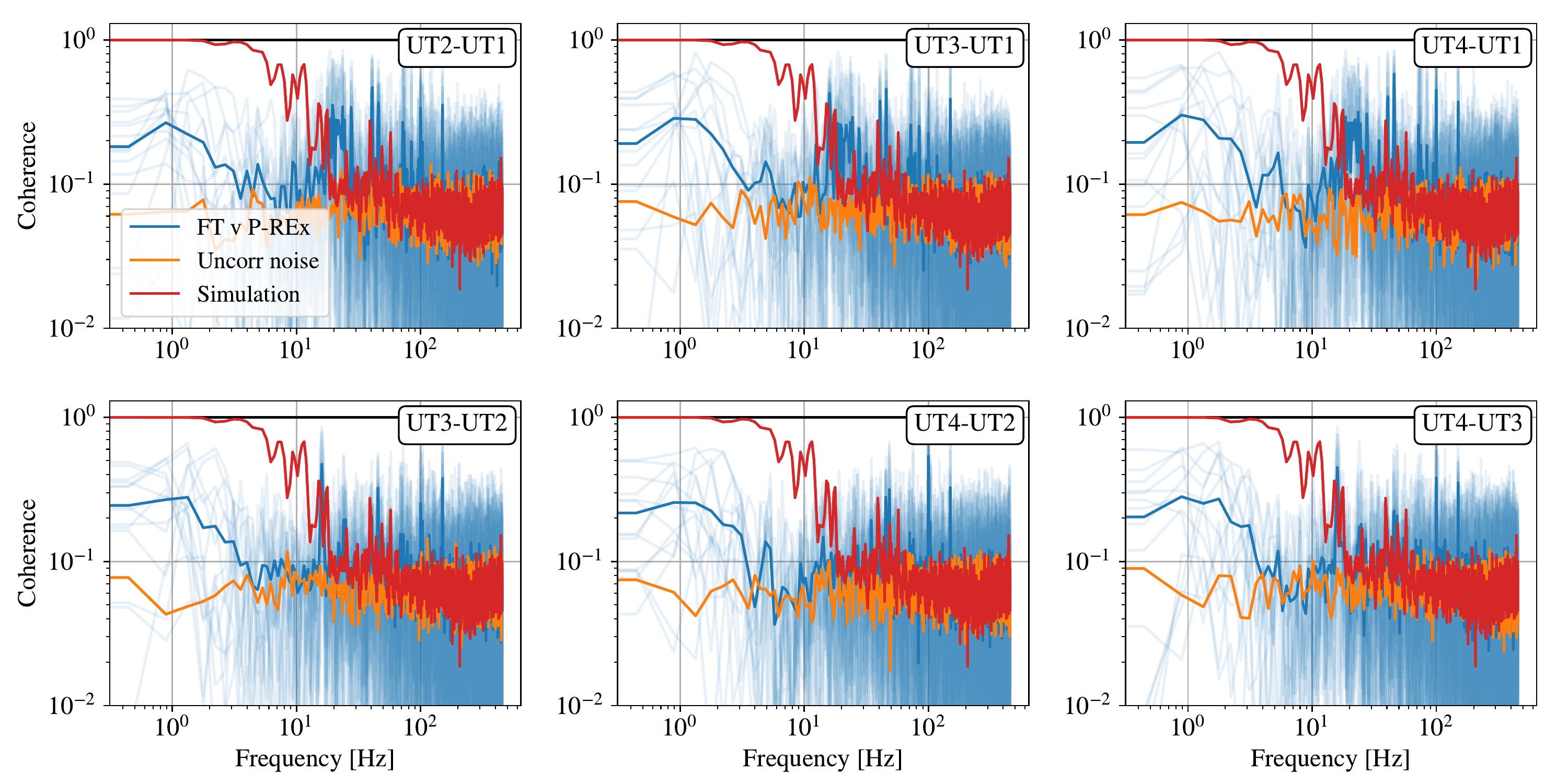}
    \caption{Average coherence of the OPD in the frequency domain of the expected performance through simulation (red), for uncorrelated signal (orange) and the coherence between the FT and PREx applied to CIAO.  }
    \label{fig:coherence-results}
\end{figure*}

Figure~\ref{fig:coherence-results} shows the coherence between the \ac{P-REx} predicted OPD and the GRAVITY FT measured OPD time series in the frequency domain. For frequencies below 4~Hz and higher frequency peaks between \mbox{10 - 100~Hz}, the obtained coherence is above the incoherent noise level. This, along with figure~\ref{fig:psd-results}, indicates that \ac{P-REx} can estimate the temporal evolution of the atmospheric OPD above the telescope, and therefore can partially identify the GRAVITY FT measured OPD. In addition, this further indicates that the FT is likely measuring an OPD in excess of that expected from atmospheric turbulence alone. The coherence at higher frequencies is possibly due to vibrations in the optical path (e.g. telescope/mirror) that produce \ac{TT} and piston, and can be observed in both the AO and FT systems. The simulated results show the coherence between the actual atmospheric OPD and the \ac{P-REx} estimated OPD for a single turbulent layer. This result indicates the potential coherence that can be achieved. The average coherence over the six baselines and 17 datasets at 1~Hz is 0.3, with some datasets exhibiting no coherence and others as high as 0.6. The figure shows that \ac{P-REx} is able to recover a high coherence for frequencies less 10~Hz. The difference between the power in the FT and \ac{P-REx} is likely due to non-atmospheric effects. This will include (i)~photon noise, indicated by the flattening in the FT PSD in figure~\ref{fig:psd-results}, (ii)~internal perturbations to the interferometer such as internal seeing, wind shake and vibrations, the latter will be confined to isolated peaks in the spectrum,  (iii)~that the curvature of the DM used by the AO system has a defocus mode very close to the piston mode, making a piston-free AO basis difficult, (iv)~the way the FT senses piston, it is possible that higher order modes such as defocus can contribute to piston aberrations \citep{Woillez19} and
(v)~the FT suffers from internal wavefront errors that offset the definition of what a flat wavefront is and therefore amplifies the high-order to piston conversion. All these effects are being investigated as part of the on-going GRAVITY+ project.

Simulated results show that for frequencies down to 1~Hz the rms can be reduced by up to a factor of 10. However, in order to utilise \ac{P-REx} the telescope and instrumental noise of the system needs to first be reduced down to or below the level of atmospheric piston drift. 

\section{Conclusions} \label{sec:conculsion}
In this paper the \ac{P-REx} algorithm and its methodology were presented and its performance capabilities were discussed in the context of simulations and post-processed observational data. The modified wind velocity algorithm was used to estimate the wind velocity of the ground layer turbulence observed by each UT's CIAO system. A strong correlation between the four CIAO systems indicated that the wind velocities did not differ significantly between the telescopes for concurrent observations. Approximately 80 $\%$ of the data agreed with the assumption that the atmosphere is dominated by a single, turbulent layer. In addition, comparisons with concurrent Paranal wind speed measurements indicate that under those conditions \ac{P-REx} can effectively estimate the wind speed.   

Comparisons between the estimated OPD values from P-REx, applied to VLTI GRAVITY CIAO WFS data, with the VLTI GRAVITY FT data showed a significant difference in power. This difference is believed to be due to telescope and instrumental noise, which is not visible to CIAO. These sources of noise will be investigated as part of the on-going GRAVITY+ project. Moving forward, until these noise contributions are brought down to the atmospheric piston noise, the \ac{P-REx} method does not have scope to improve on the VLTI performance or sensitivity. However, comparisons of \ac{P-REx} applied to simulated data and real on-sky data indicate that \ac{P-REx} can accurately estimate the atmospheric OPD. Therefore, if the telescope and instrumental noise of the VLTI GRAVITY instrument is suppressed, \ac{P-REx} has the potential to reduce the atmospheric OPD \ac{rms} by up to a factor of 10.

A potential next step would be to test \ac{P-REx} on data taken by the \ac{LBTI}, which is a simpler, single telescope mount Fizeau interferometer system that does not employ long delay lines, and therefore is expected to reduce the opto-mechanical OPD vibrations. Additionally, the LBTI already employs active star-light free vibration control in the telescope \citep{Bohm16, Bohm17}, which reduces the vibration level and increases the signal-to-noise of the atmospheric turbulence. This will result in a better prediction performance by P-REx.

\section*{Acknowledgements}
This work was funded by WP8 of OPTICON/H2020 (2017-2020), grant agreement 730890 (https://www.astro-opticon.org/). We would like to thank Stefan Hippler for communications with regards to the CIAO data and Robert J. Harris for his general support throughout this project. The authors would also like to thank the reviewer, Dr Richard W. Wilson, for their comments and efforts towards improving this manuscript.

\section*{Data Availability}
The data in this underlying article was provided by the GRAVITY consortium by permission. Data will be shared on request to the corresponding author with permission of the GRAVITY consortium. 



\bibliographystyle{mnras}
\bibliography{ref}








\bsp	
\label{lastpage}
\end{document}